\documentclass[reprint,amsmath,amssymb,aps,showpacs,pra]{revtex4-1}
\usepackage{graphicx}
\usepackage{dcolumn}
\usepackage{bm}
\usepackage[usenames,dvipsnames]{color}
\usepackage{caption}
\usepackage{subcaption}
\usepackage[utf8]{inputenc}
\usepackage{hyperref}
\graphicspath{{./Files/PDFs/}}

\begin{document}


\title{Recurrence in three dimensional magnetohydrodynamic plasma}

\author{Rupak Mukherjee}
\email{rupakmukherjee01@gmail.com; rupak@ipr.res.in}
\author{Rajaraman Ganesh}
\author{Abhijit Sen}
\email{abhijit@ipr.res.in}
\affiliation{Institute for Plasma Research, HBNI, Bhat, Gandhinagar - 382428, India}

\begin{abstract}
We report on a numerical observation of recurrence phenomenon in a three dimensional magneto-hydrodynamic (MHD) plasma for certain classes of initial flow profiles. Our simulations discover such a behaviour for an initial Taylor-Green (TG) type of flow whereas under identical conditions an initial Arnold-Beltrami-Childress (ABC) flow fails to show recurrent behaviour. This difference in the dynamical behaviours is traced to the nature of the Rayleigh quotient in each case. The Rayleigh quotient which is an approximate measure of the number of active degrees of freedom in a dynamical system is found  to remain small and bounded in time for the TG flow but grows indefinitely in time for the ABC flow. An estimate of the effective degrees of freedom for the TG flow case is provided and its practical implications for the nonlinear behaviour of the continuum nonlinear MHD plasma system is discussed. 
\end{abstract}

\maketitle


\section{Introduction}
\label{sec:intro} 
Recurrence phenomenon which implies a repeated occurence of a `state' in a nonlinear dynamical system is an intriguing and counterintuitive feature of many body systems that have attracted a great deal of past attention from physicists and mathematicians. One of the most celebrated examples of recurrence is the Fermi-Pasta-Ulam-Tsingtou (FPUT) problem  \cite{dauxois:2008} in which the initial energy put into a single mode of a one dimensional chain of nonlinear oscillators did not lead to a thermalization of the system but saw the energy periodically returning to the initial state thereby invalidating the existence of erogidicity in nonlinear systems. Attempts to explain the FPUT `puzzle' uncovered fascinating underlying links between the phenomenon and integrable equations possessing soliton solutions \cite{zabusky:1965} and with the Henon-Heiles problem that exhibits chaos \cite{chirikov:1979}. An FPUT like recurrence was subsequently observed in a two dimensional nonlinear Schrodinger equation by Yeun and Ferguson \cite{yuen:1978}.  
Thyagaraja \cite{thyagaraja:1979} provided a simple analytic explanation of their results by arguing that many nonlinear systems can be shown to have a finite number of effective degrees of freedom even though the evolution equations belong to a continuum. He also provided an explicit estimate of the effective degrees of freedom for the nonlinear Schrodinger equation model. He further pointed out the possibility of the Navier-Stokes equation possessing a finite number of effective degrees of freedom in the case of the Benard-Couette flow thereby extending a conjecture by Landau \cite{landau:1959} who had predicted a quasi-periodic motion. While in principle, following the pioneering work of Birkhoff \cite{birkhoff:1960}, for systems with finitely many degrees of freedom, almost all motions are recurrent; in practice the criteria for observing a non-collapsing recurrent dynamics in higher dimensional systems becomes quite complicated and no general criteria have been found in the Lagrange and Poisson stability approaches \cite{thyagaraja:1981,thyagaraja:1983}. Experimentally recurrence phenomena have been  observed in shallow water waves \cite{elgar:1990}, ocean waves \cite{bryant:1988}, Couette turbulence \cite{viswanath:2007} and in quantum dynamics \cite{hogg:1982}. A detailed description of recurrence in dynamical systems in the sense of Poincare recurrence has been given by Katok and Hasselblatt \cite{katok:1997}.

In plasmas, Tajima {\it et al} \cite{tajima:1981} had observed a complete reconstruction of Langmuir wave packets following its initial break-up in two spatial dimensional electrostatic particle simulation including ion and electron dynamics. Kaw {\it et al} \cite{kaw:1973} identified such reconstruction as originating from the spatial density dependence of the plasma frequency that causes in and out of phase oscillations. In principle, all periodic or quasiperiodic motion can be considered as examples of ``recurrent dynamics''. Such behavior is commonly found in many nonlinear driven dissipative systems. Sawteeth, ELMs and many similar phenomena in tokamaks, are prime examples of such phenomena that have been well studied numerically in this context \cite{vlad:1989,thyagaraja:1999,pamela:2011,jardin:2012}. Periodicity results if essentially two degrees of freedom are mainly ``active" whilst quasi periodicity involves a few incommensurate frequencies. Truly recurrent motions however occur when the recurrence time is long compared with typical frequencies of the system. It is such behavior found in nonlinear continuum systems that is at the heart of past studies of recurrence and is also the focus of our present investigation.

To the best of our knowledge a complete recurrence of chaotic kinetic and magnetic field structure has never been observed in electromagnetic simulations of magnetised plasmas. In this paper we report a first observation of recurrence in numerical simulations of the three dimensional magneto-hydrodynamic equations. The recurrence is observed for certain initial flow profiles  
and not for others. In particular we find that an initial Taylor-Green type flow profile repeats itself after a certain interval despite undergoing complex distortions and convolutions in intermediate time frames. An Arnold-Beltrami-Childress flow profile, on the other hand, launched under identical conditions does not show any recurrent behaviour. We explain our results by invoking the existence of a finite number of effective degrees of freedom put forward by Thyagaraja \cite{thyagaraja:1979}. A quantitative measure of the number of effective degrees of freedom is provided by the Rayleigh quotient defined later in the paper. We find that the Rayleigh quotient has a small finite value and remains bounded in time for the TG flow case whereas it shows an unbounded increase in time for the ABC flow case.\\
The paper is organized as follows. 
The governing equations of MHD plasma, whose numerical results are analysed, are explicitly provided in the next section. Section \ref{sec:code} briefly describes the details of our code MHD3D. The initial profiles of density, velocity and magnetic field, the boundary conditions and the parameters for which the code is run are mentioned in section \ref{sec:IC}. Section \ref{sec:results} deals with the numerical results obtained from the MHD3D code for two different initial flow profiles. A detailed analysis of the numerical results are provided in section \ref{sec:analytical} and the flow profiles leading to recurrence phenomenon are isolated. Section \ref{sec:summary} summarizes our results and analysis and discusses some of the limitations of the present work along with  suggestions for future directions of study. 


\section{Governing Equations} 
\label{sec:equn}
The basic equations governing the dynamics of the single fluid MHD plasma are as follows: 
\begin{eqnarray}
&& \label{density} \frac{\partial \rho}{\partial t} + \vec{\nabla} \cdot \left(\rho \vec{u}\right) = 0\\
&& \frac{\partial (\rho \vec{u})}{\partial t} + \vec{\nabla} \cdot \left[ \rho \vec{u} \otimes \vec{u} + \left(P + \frac{B^2}{2}\right){\bf{I}} - \vec{B}\otimes\vec{B} \right]\nonumber \\
&& \label{velocity} ~~~~~~~~~ = \mu \nabla^2 \vec{u}\\
&& P = C_s^2 \rho \\
&& \label{Bfield} \frac{\partial \vec{B}}{\partial t} + \vec{\nabla} \cdot \left( \vec{u} \otimes \vec{B} - \vec{B} \otimes \vec{u}\right) = \eta \nabla^2 \vec{B}
\end{eqnarray}
In this system of equations, $\rho$, $\vec{u}$, $P$ and $\vec{B}$ are the density, velocity, kinetic pressure and magnetic field of a fluid element respectively. $\mu$ and $\eta$ denote the coefficient of kinematic viscosity and magnetic resistivity. We assume $\mu$ and $\eta$ are constants over space and time. The symbol ``$\otimes$'' represents the dyadic between two vector quantities. 

The kinetic Reynold's number ($Re$) and magnetic Reynold's number ($Rm$) are defined as $Re = \frac{U_0 L}{\mu}$ and $Rm = \frac{U_0 L}{\eta}$ where $U_0$ is the maximum velocity of the fluid system to start with and $L$ is the system length. 

We also define the sound speed of the fluid as $C_s = \frac{U_0}{M_s}$, where, $M_s$ is the sonic Mach number of the fluid. The Alfven speed is determined from $V_A = \frac{U_0}{M_A}$ where $M_A$ is the Alfven Mach number of the plasma. The initial magnetic field present in the plasma is determined from the relation $B_0 = V_A \sqrt{\rho_0}$, where, $\rho_0$ is the initial density profile of the fluid.
   

\section{Numerical Method}
\label{sec:code}
In order to simulate the plasma dynamics governed by the above-mentioned MHD equations, a direct numerical simulation (DNS) code, MHD3D, has been developed in-house at the Institute for Plasma Research. MHD3D is an OpenMP parallel three dimensional weakly compressible, viscous, resistive magnetohydrodynamic code that uses a pseudo-spectral technique to simulate a general scenario of three dimensional magnetohydrodynamic turbulence in a Cartesian box with periodic boundary conditions. The pseudo-spectral technique, one of the most accurate computational fluid dynamic (CFD) techniques available today, uses the FFTW libraries \cite{FFTW3:2005} which is one of the fastest Fourier Transform libraries developed recently. This technique is applied to calculate the spatial derivatives and to evaluate the non-linear terms involved in the model equations with a standard de-aliasing using the 2/3 truncation rule. The time derivative is solved using multiple explicit time solvers viz. Adams-Bashforth, Runge-Kutta 4 and Predictor-Correcter algorithms and all the solvers have been checked to provide identical results. The divergence of the magnetic field is found to be ${\cal{O}} (10^{-35})$ till the end of the simulation.

The three dimensional weakly compressible neutral fluid solver is benchmarked with the results of Samtaney {\it et al}\cite{ravi:2001} for the root mean square of velocity divergence and the skewness of velocity field for a decaying turbulence case. The three dimensional kinetic dynamo effect is matched with Galloway {\it et al} \cite{galloway:1986} in the case of ABC flow for a grid resolution of $64^3$. For iso-surface plots we use the open-source 3D scientific data visualization and plotting application of python named ``{\it{MayaVi}}'' \cite{mayavi}. Further benchmarking details and protocols followed in the code can be found in earlier works \cite{rupak:2018a,rupak:2018b,rupak:2018c}.

\begin{figure*}[h!]
\includegraphics[scale=0.65]{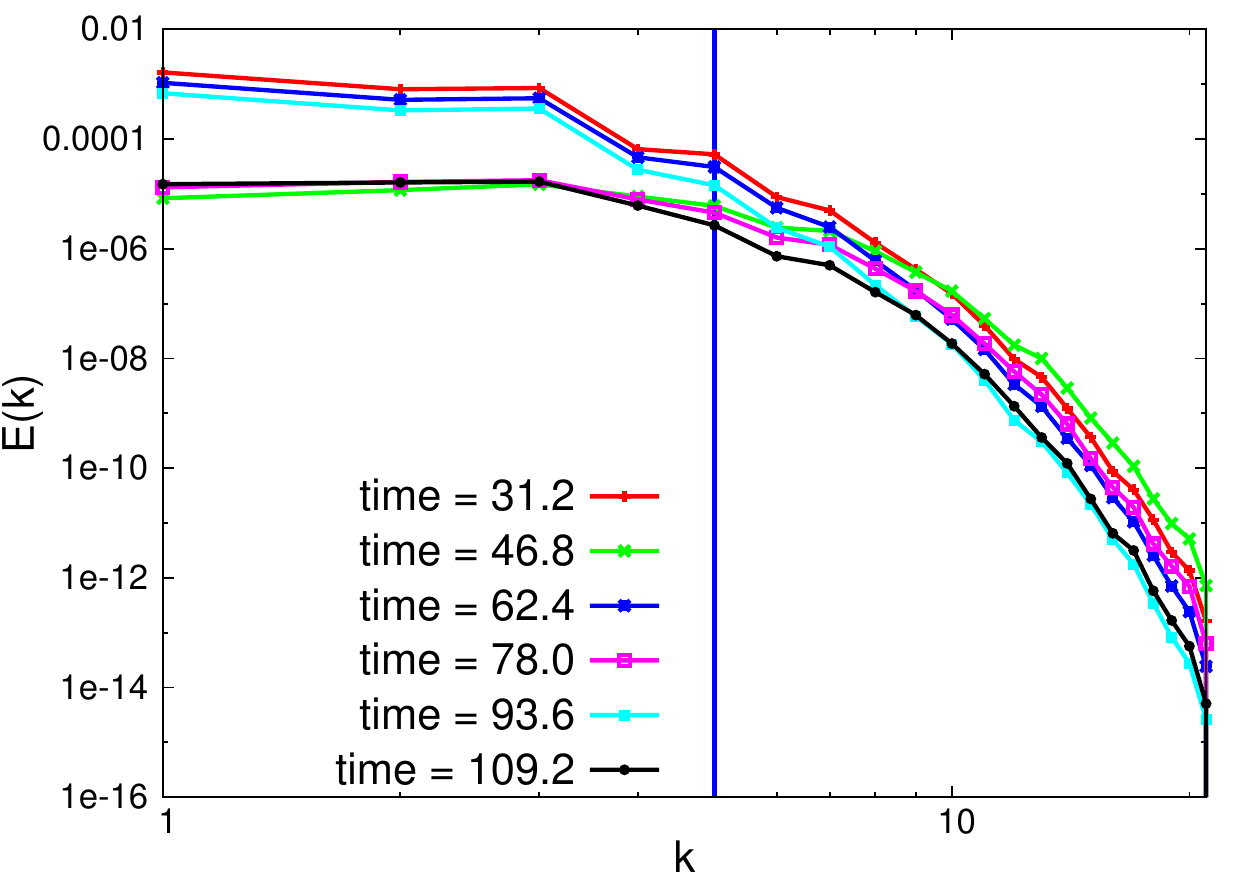}
\includegraphics[scale=0.65]{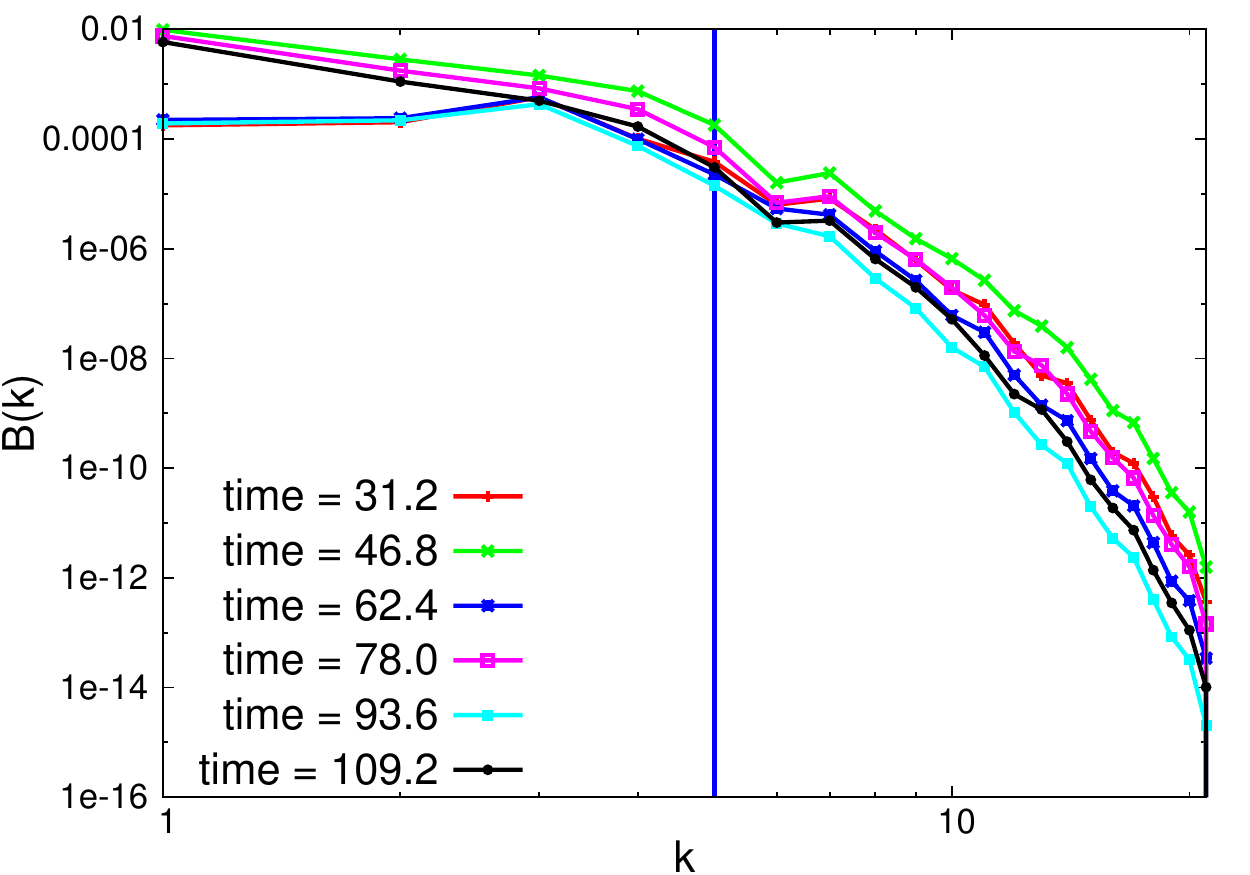}
\caption{(Color online) The kinetic ($E_k$ vs $k$) and magnetic ($B_k$ vs $k$) energy spectra from long time evolution of the Taylor-Green velocity profile. Even after several recurrence cycles, the energies are found to be contained in the large spatial scales only. The oscillation of the first five modes with time indicates a recurrence phenomenon. The change in amplitude of the energy oscillation (e.g. in the fundamental mode) spans two orders of magnitude.}
\label{TG_Spectra}
\end{figure*}

To check whether a $64^3$ resolution is sufficient for the current study, we plotted the energy spectra for the Taylor-Green flow and found that most of the energy was located in the first $5$ modes [Fig.\ref{TG_Spectra}]. This is in accord with earlier higher resolution studies and indicates that the present resolution is adequate for our simulation studies of recurrence. We have repeated the same diagnostics test for the Arnold-Beltram-Childress flow and found similar spectra containing energy within the first $5$ modes. In Fig.[\ref{TG_Spectra}], the kinetic energy spectra ($E_k = \sum \frac{u_k^2}{2}$ vs $k$) is plotted at times $t = 31.2, 62.4, 93.6$ which correspond to the recurrence times or coherent oscillation period of the kinetic energy. We also plot the energy spectra at times halfway between the recurrence times i.e. at $t = 46.8, 78.0, 109.2$. Also, the magnetic energy spectra ($B_k = \sum \frac{B_k^2}{2}$ vs $k$) is seen to oscillate in an anti-phase manner with the kinetic energy spectra [Fig. \ref{TG_Spectra}]. It is interesting to note that the change in amplitude of the energy oscillation (e.g. in the fundamental mode) spans two orders of magnitude.\\


\section{Initial and boundary conditions}
\label{sec:IC}
The choice of an appropriate initial velocity profile is very crucial for the observation of recurrence of the initial flow structure. The identification of the class of velocity profiles giving rise to recurrence phenomenon is described at length in Section \ref{sec:analytical}. We take two example flows, one showing bounded Rayleigh quotient and the other having unbounded Rayleigh quotient and perform our numerical analysis. The flows we take up, are quite well known and well discussed in past literature. The first one we consider is the Taylor-Green (TG) flow for which we report the recurrence phenomenon. The other one is the Arnold-Beltrami-Childress (ABC) flow which shows the absence of recurrence, as expected from our analytical insight discussed later.

The initial density profile is chosen to be uniform throughout the simulation domain. The initial magnetic field profile is chosen as $B_x = B_y = B_z = B_0$ throughout the space. Our simulation domain is periodic in all the three spatial directions and thus may be considered to represent a small part of an infinitely large plasma system.  


\subsection{Parameter details}
\label{subsec:parameter}
The parameters for which we run our code MHD3D are given in Table \ref{parameter}.
\begin{table}[h!]
\centering
\begin{tabular}{ |c|c|c|c|c|c|c|c|c| }
 \hline
 $N$ & $L$ & $dt$ & $\rho_0$ & $U_0$ & $Re$ & $Rm$ & $M_s$ & $M_A$ \\
 \hline
 ~ & ~ & ~ & ~ & ~ & ~ & ~ & ~ &\\
 $64$ & $2 \pi$ & $10^{-5}$ & $1$ & $0.1$ & $450$ & $450$ & $0.1$ & $1$ \\
 ~ & ~ & ~ & ~ & ~ & ~ & ~ & ~ &\\ 
 \hline
\end{tabular}
\caption{Parameter details for the results mentioned in this report. These parameters are kept identical throughout this report unless stated otherwise.}
\label{parameter}
\end{table}
The OpenMP parallel MHD3D code is run on $20$ cores for $9600$ CPU hours for a single set of parameters. From our energy spectra calculation we find that most of the energy resides in the large scales and hence a grid resolution of $64^3$ is believed to be good enough to resolve the phenomena of interest. In some test runs, we have increased the grid resolution to $128^3$ and found no significant variation from the results of $64^3$. Similarly a single run with grid resolution $64^3$ and time stepping width $dt = 10^{-6}$ produced identical result to that of  $dt = 10^{-5}$. The magnitude of initial density is known to affect the dynamics of compressible fluid \cite{bayly:1992, terakado:2014}. However, the effect of variation of the magnitude of initial density is beyond the scope of this present work and will be addressed in a future communication.  


\section{3D DNS Results from MHD3D}
\label{sec:results}
We time evolve the above-mentioned initial density, velocity and magnetic field profiles according to the equations mentioned in Section \ref{sec:equn} using our DNS code - MHD3D and obtain the following results. 


\subsection{Taylor-Green flow}
\label{subsec:TG}
\begin{figure*}[h!]
\includegraphics[scale=0.65]{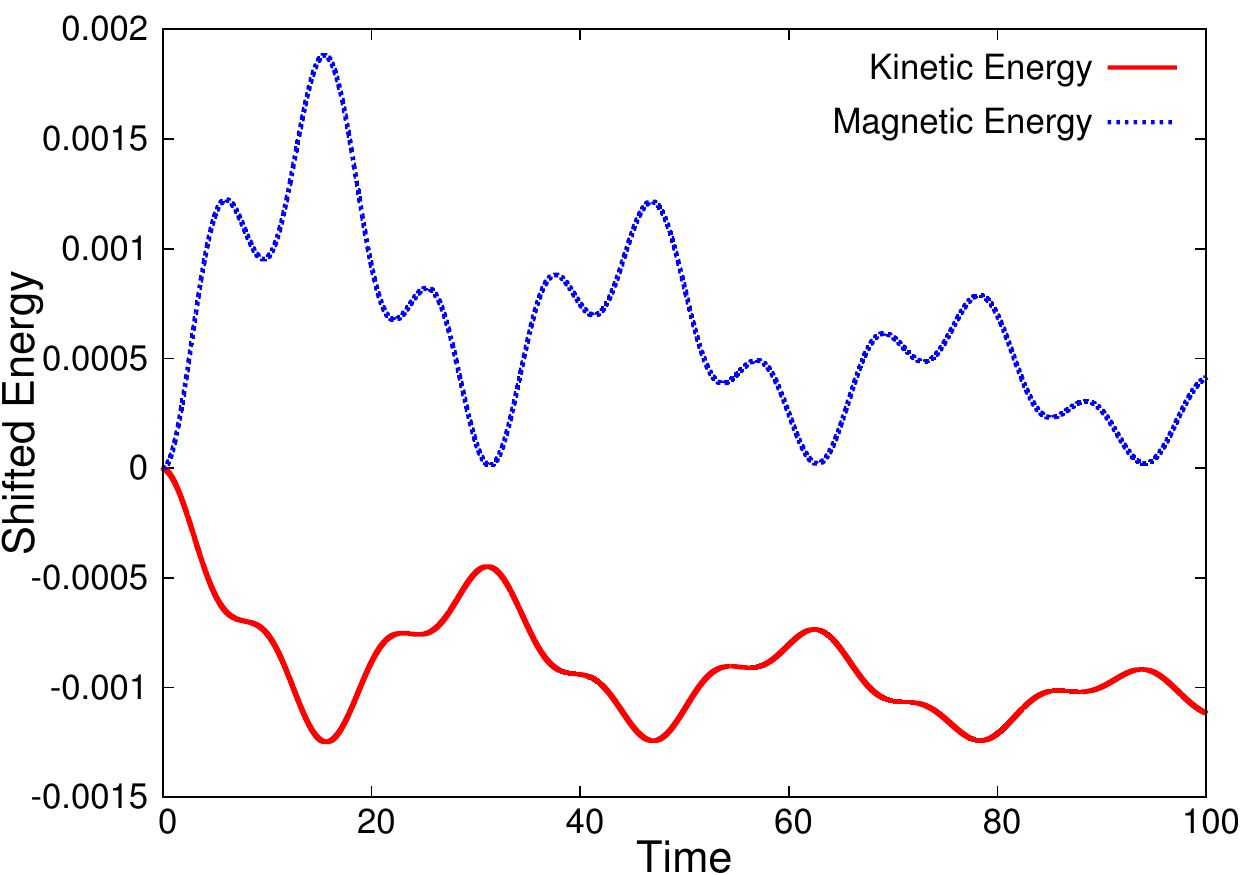}
\caption{(Color online) The shifted kinetic [$\vec{u}^2(t) - \vec{u}^2(0)$] (red solid) and magnetic [$\vec{B}^2(t) - \vec{B}^2(0)$] (blue dotted) energies of Taylor - Green flow oscillate with time exchanging energy between their respective modes. The shift is measured by the removal of the initial magnitude of kinetic and magnetic energy respectively.}
\label{TG_Energy}
\end{figure*}

The velocity profile of MHD plasma subjected to divergence-free Taylor-Green (TG) flow is 
\begin{equation}\label{TG}
\begin{aligned}
u_x &= A ~ U_0 \left[ \cos(kx) \sin(ky) \cos(kz) \right]\\
u_y &= - A ~ U_0 \left[ \sin(kx) \cos(ky) \cos(kz) \right]\\
u_z &= 0
\end{aligned}
\end{equation}
We choose $A = 1$ and $k = 1$ in this study. The kinetic and magnetic energies oscillate as a result of continuous conversion and exchange of energy between the two modes (Fig. \ref{TG_Energy}). The decay in the total energy is solely due to viscous and resistive effects.\\

\begin{figure*}[h!]
\includegraphics[height=25cm,width=20cm]{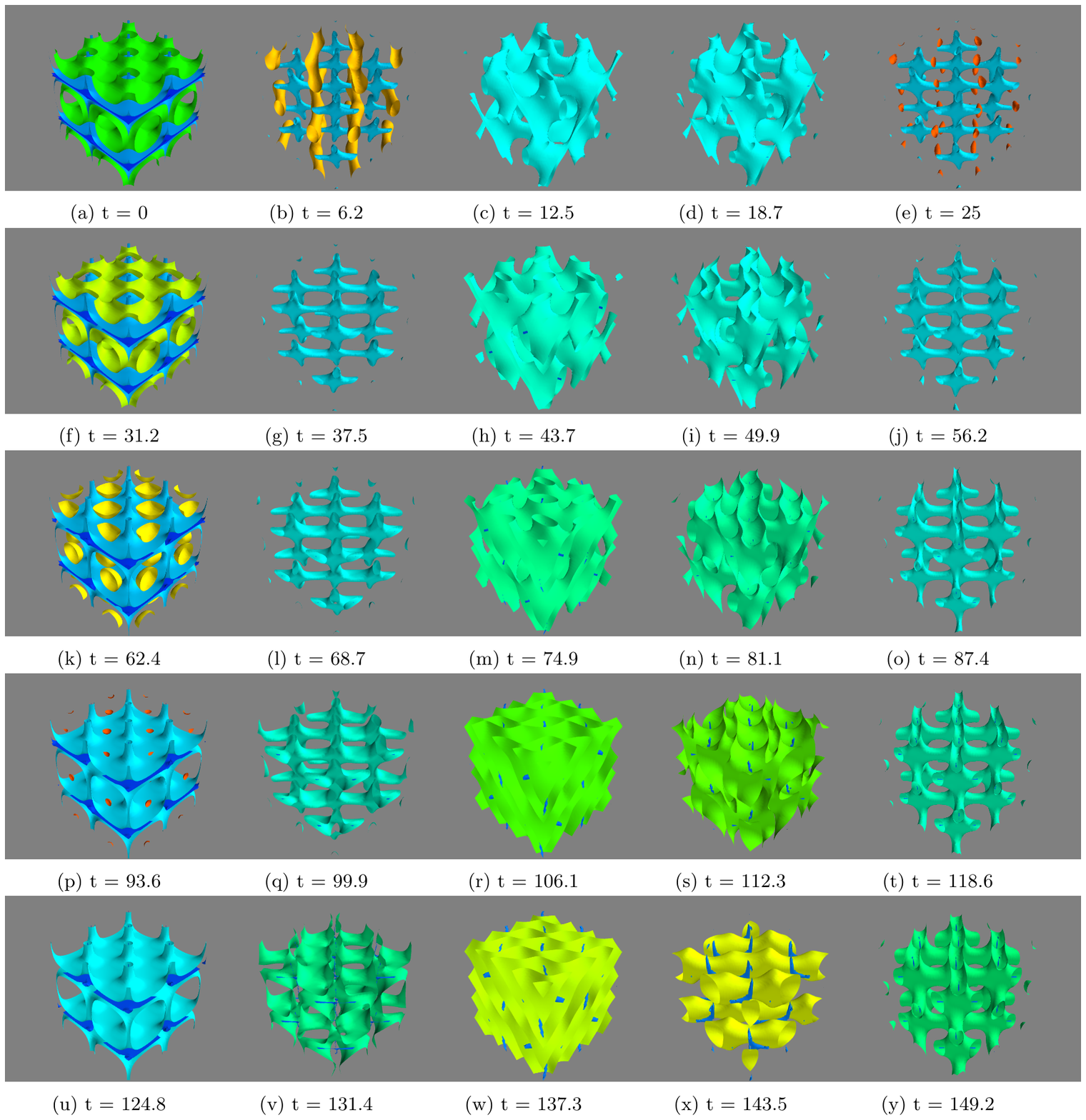}
\vspace{-6cm}
\caption{Recurrence of velocity field isosurfaces for Taylor-Green flow [Eq. \ref{TG}] for the magnitudes $0.001$, $0.05$, $0.01$. The initial velocity field isosurfaces (at time $t = 0$) are found to recur at time $t = 31.2, 62.4, 93.6, 124.8$. The slight deviation from the initial isosurface is solely due to the viscous and resistive effects present in the governing dynamics equation for regularisation.}
\label{TG_vel_iso}
\end{figure*}

\begin{figure*}[t!]
\includegraphics[height=25cm,width=20cm]{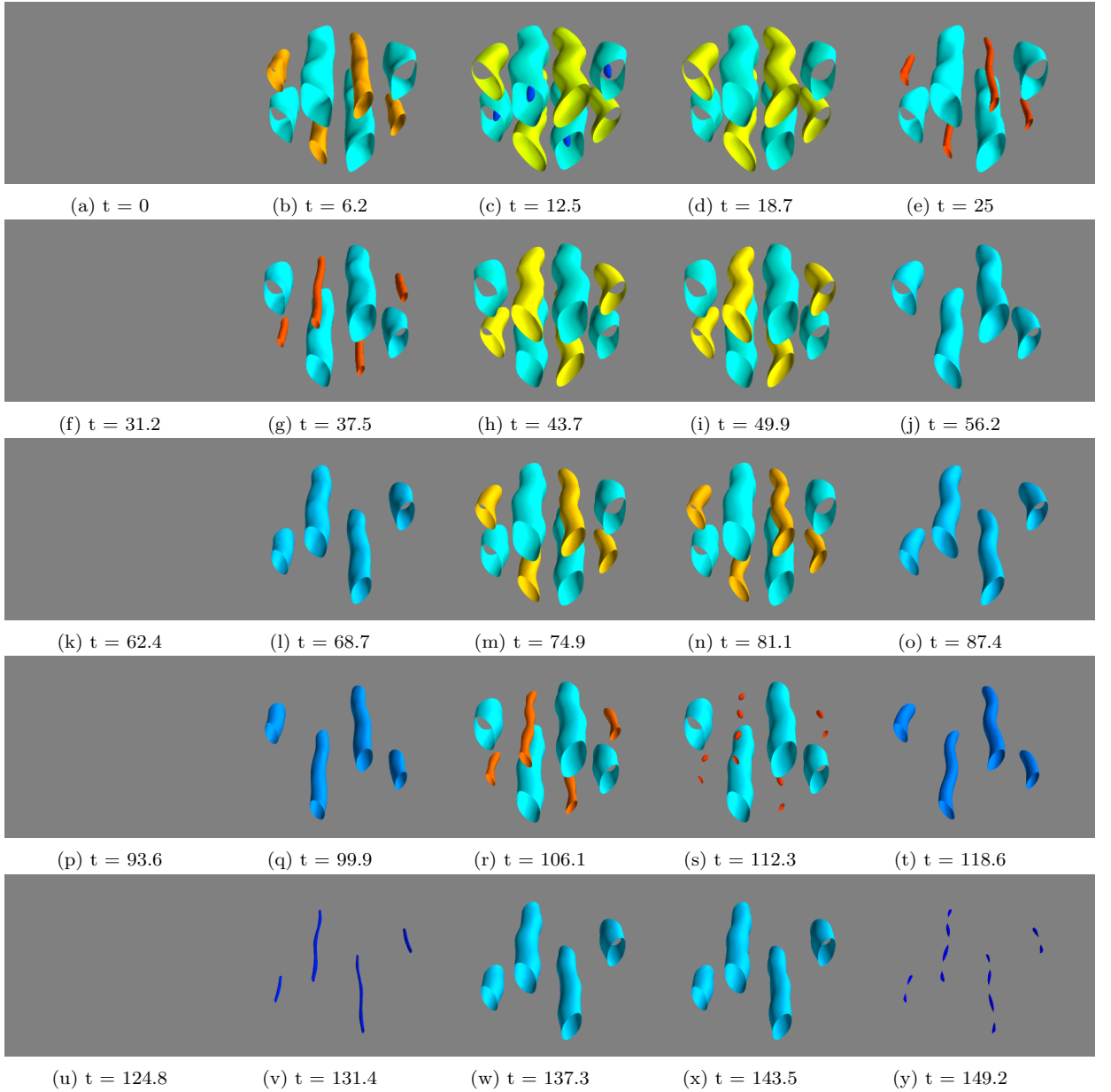}
\vspace{-6cm}
\caption{Recurrence of magnetic field isosurfaces for Taylor-Green flow for the magnitudes $0.13$, $0.16$, $0.2$. The initial magnetic field isosurfaces (at time $t = 0$) are found to recur at time $t = 31.2, 62.4, 93.6, 124.8$. The slight deviation from the initial isosurface is solely due to the viscous and resistive effects present in the governing dynamics equation for regularisation.}
\label{TG_mag_iso}
\end{figure*}

The velocity and magnetic field isosurfaces are plotted in Fig \ref{TG_vel_iso} and \ref{TG_mag_iso} respectively. In Fig. \ref{TG_vel_iso} the blue, sky and green isosurfaces are plotted for the values $0.001$, $0.05$, $0.01$ and similarly in Fig. \ref{TG_mag_iso} for $0.13$, $0.16$, $0.2$ respectively. From Fig. \ref{TG_vel_iso} we observe that the flow profile is repeated after a time interval of about $t = 31.2$. It can be easily noted that in Fig. \ref{TG_vel_iso} [(a), (f) and (k)] show similar flow profiles. Similarly, [(b), (g), (l)], [(c), (h), (m)], [(d), (i), (n)] and [(e), (j), (o)] show similar flow structures though some of the isosurfaces present in (b) or (e) do not appear in (g) or (j) and (l) or (o) because of viscous and resistive effects. After the third cycle, the initial condition is not exactly reproduced. However the structures of the flow profile continue to remain same. The repeated occurrence of similar profiles represents a ``recurrence'' phenomenon. Similar to velocity profiles, magnetic field profiles also show a recurrence phenomenon. At $t = 0$ the magnetic field is uniform and hence, no flow structure remains, thus no isosurface is observed for several distinct values of magnitudes mentioned above. The uniform field profile comes back after the same time $t = 31.2$ as observed for velocity profiles. Identical to velocity profiles in Fig. \ref{TG_mag_iso} the same corresponding sub-figures show similar profiles of isosurfaces after time intervals of $31.2$ unit. \\ 

\begin{figure*}[h!]
\includegraphics[scale=0.65]{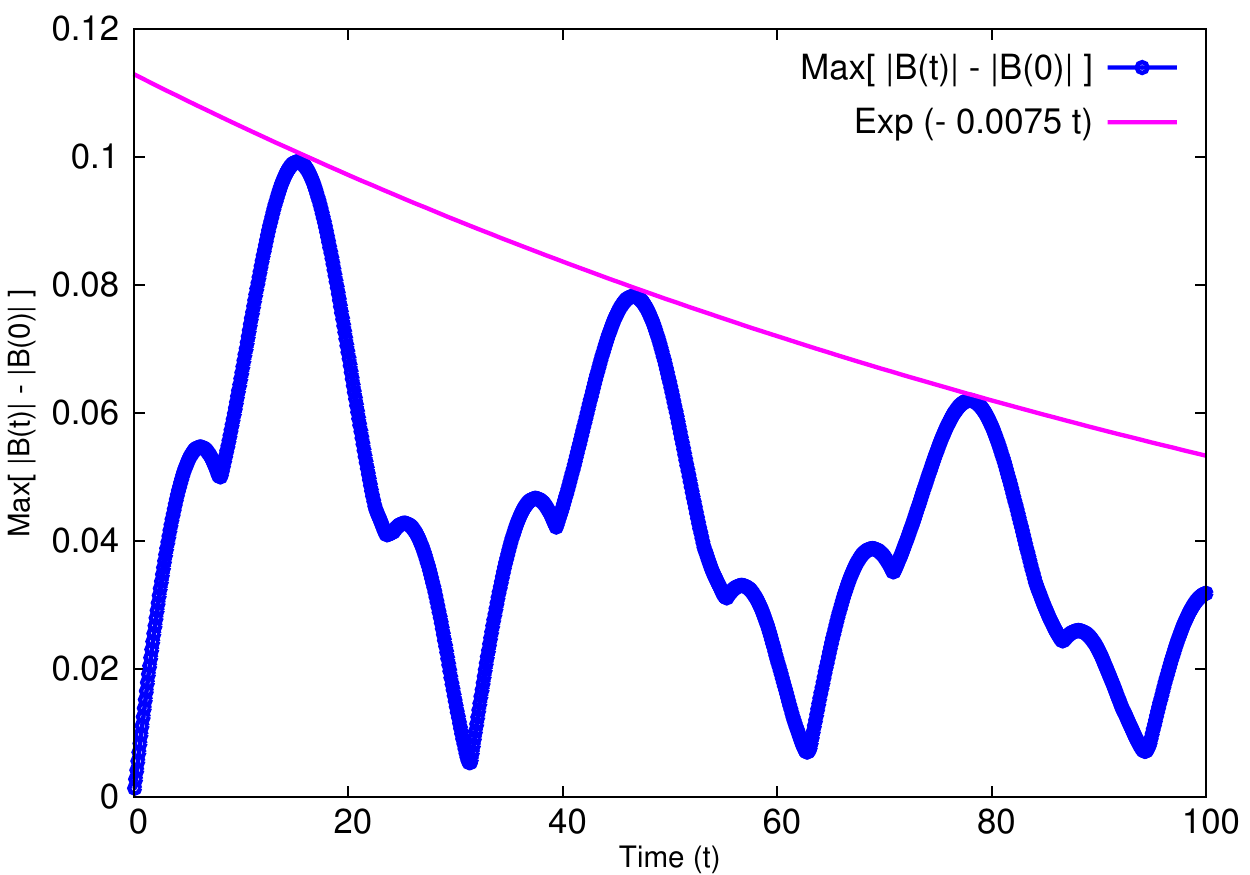}
\caption{(Color online) The time evolution of $f = Max [|B(t)| - |B(0)|]$ for Taylor-Green (TG) flow showing that $f$ regularly reaches a minimum (nearly zero) at times $31.2$, $62.4$ and $93.6$ respectively. The decay in the amplitude of the oscillatory behavior is on account of dissipation in the system arising from finite viscosity and resistivity contributions.}
\label{Referee}
\end{figure*}

To provide a more quantitative assessment of the observed qualitative recurrence phenomenon we have taken the difference in the values of the magnetic field at successive times after the initial time and examined the evolution of this difference quantity. We define a quantity $f = Max [|B(t)| - |B(0)|]$ and show its time evolution for three successive recurrence cycles in Fig. ~\ref{Referee}. As can be seen, the quantity $f$ regularly reaches a minimum (nearly zero) at times $31.2$, $62.4$ and $93.6$ thus providing a clear and quantitative signature of the existence of recurrence. We also note that the peak amplitude of $f$ decays in time. This is a consequence of the existence of dissipation in the system due to the finite values of the viscosity and resistivity coefficients. An exponential curve is fitted connecting the peaks of the quantity $f$ to highlight this dissipative decay.\\

Since the kinetic and magnetic energy primarily reside in the large spatial scales of the spectrum (Fig.\ref{TG_Spectra}) a better grid resolution with a smaller time-stepping will not change the fundamental nature of our above results. However lesser values of $\mu$ and $\eta$ might show a larger number of repetitive cycles, though the computational cost of the runs to resolve the small length-scales will be much higher than the present runs.


\subsection{Arnold-Beltrami-Childress flow}
\label{subsec:ABC}
\begin{figure*}[h!]
\includegraphics[scale=0.65]{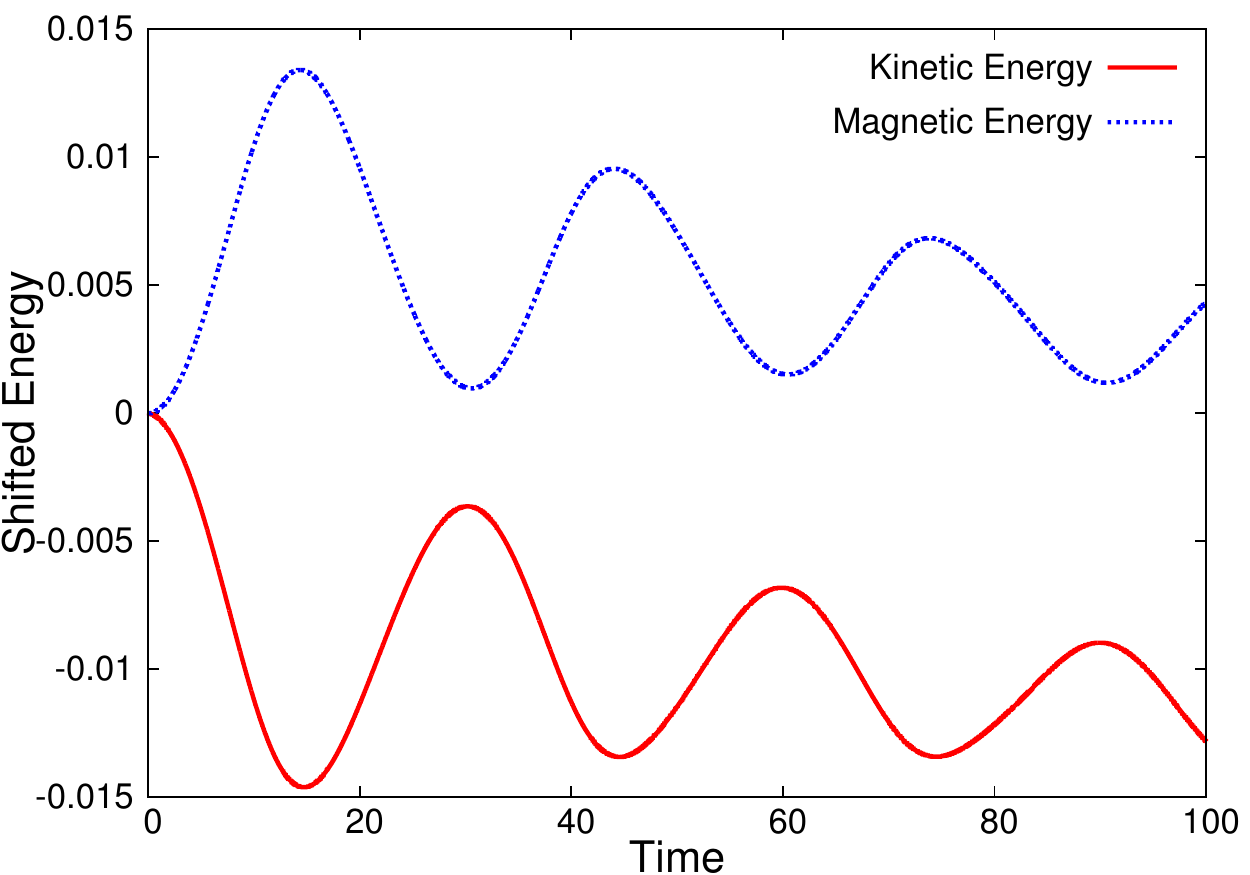}
\caption{(Color online) The shifted kinetic (red solid) and magnetic (blue dotted) energies for Arnold-Beltrami-Childress flow oscillate with time exchanging energy between their respective modes. The shift is measured by the removal of the initial magnitude of kinetic and magnetic energy respectively.}
\label{ABC_Energy}
\end{figure*}

We choose another velocity profile of MHD plasma subjected to divergence-free Arnold-Beltrami-Childress (ABC) flow as 
\begin{equation}\label{ABC}
\begin{aligned}
u_x &= U_0 [ A \sin(kz) + C \cos(ky) ]\\
u_y &= U_0 [ B \sin(kx) + A \cos(kz) ]\\
u_z &= U_0 [ C \sin(ky) + B \cos(kx) ]
\end{aligned}
\end{equation}
$A = B = C = 1$ and $k = 1$ are chosen for this study. The kinetic and magnetic energies oscillate as a result of continuous conversion and exchange of energy between the two modes (Fig. \ref{ABC_Energy}). The decay in the total energy is once again solely due to viscous and resistive effects.\\

\begin{figure*}[h!]
\includegraphics[height=25cm,width=20cm]{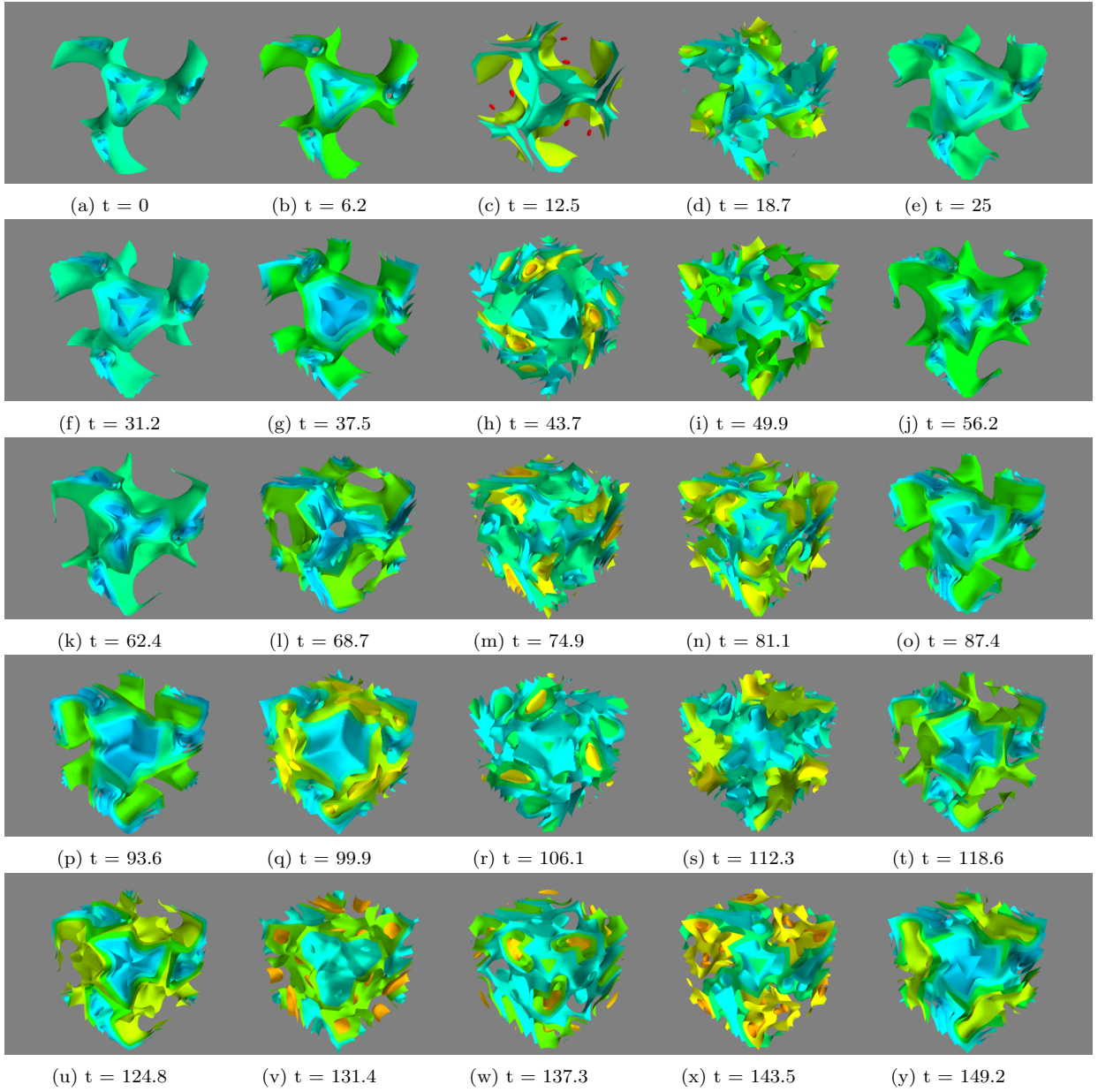}
\vspace{-6cm}
\caption{Time evolution of velocity field isosurfaces for Arnold-Beltrami-Childress flow [Eq.\ref{ABC}] for the magnitudes $0.03$, $0.05$, $0.08$, $0.1$. Significant deviation is obtained from the initial kvelocity field isosurfaces (at time $t = 0$) at time $t = 31.2, 62.4, 93.6, 124.8$ showing the absence of recurrence phenomena.}
\label{ABC_vel_iso}
\end{figure*}

\begin{figure*}[h!]
\includegraphics[height=25cm,width=20cm]{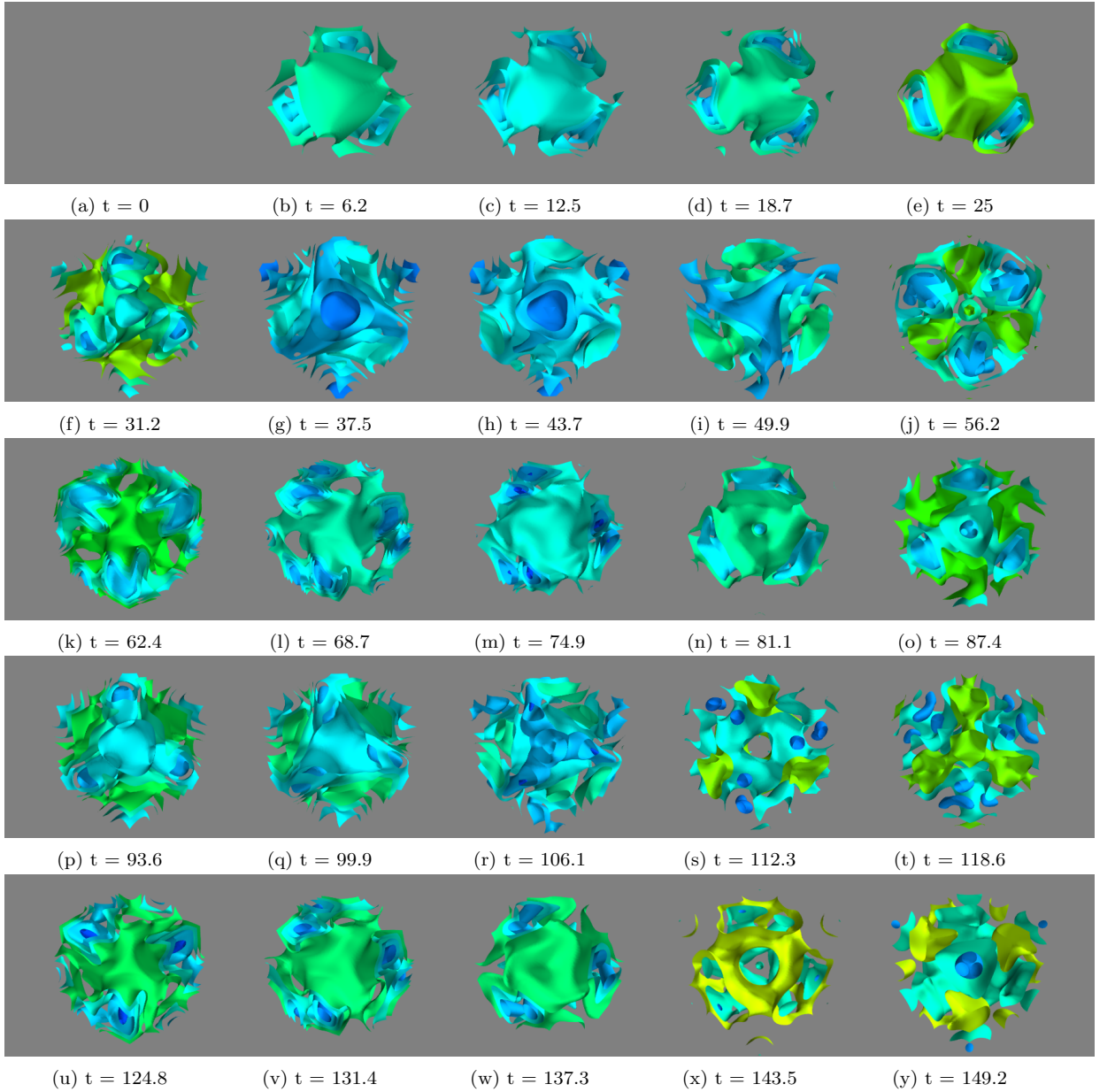}
\vspace{-6cm}
\caption{Time evolution of magnetic field isosurfaces for Arnold-Beltrami-Childress flow for the magnitudes $0.1$, $0.133$, $0.166$, $0.2$. Significant deviation is obtained from the initial magnetic field isosurfaces (at time $t = 0$) at time $t = 31.2, 62.4, 93.6, 124.8$ showing the absence of recurrence phenomena.}
\label{ABC_mag_iso}
\end{figure*}

The velocity and magnetic field isosurfaces are plotted in Fig. \ref{ABC_vel_iso} and \ref{ABC_mag_iso} respectively. In Fig. \ref{ABC_vel_iso} the blue, sky, yellow and green isosurfaces are plotted for the values $0.03$, $0.05$, $0.08$, $0.1$ and similarly in Fig. \ref{ABC_mag_iso} for $0.1$, $0.133$, $0.166$, $0.2$ respectively. As can be seen, unlike the TG flow case, the velocity profile is unable to reconstruct the initial state closely at time $t = 31.2$ or subsequently.  This deviation is not due to viscous and resistive effects and that can be verified from the profiles of other time instants for example in Fig. \ref{ABC_vel_iso} subfigures [(b), (g), (l), (q), (v)], [(c), (h), (m), (r), (w)], [(d), (i), (n), (s), (x)] and [(e), (j), (o), (t), (y)]. The significant deviation can be well observed from the magnetic field profiles also as shown in Fig. \ref{ABC_mag_iso}\\


\section{Analytical description of the DNS results}
\label{sec:analytical}

The numerical results in section \ref{sec:results} show that, there exists a certain class of flows that periodically reproduces the initial flow profile (Fig. \ref{TG_vel_iso}). Simultaneously the magnetic field structures are also reproduced in cycles (Fig. \ref{TG_mag_iso}). In order to gain a physical understanding of this phenomena we examine the Rayleigh Quotient which effectively represents the ``active number of degrees of freedom'' of a system. \\

\begin{figure*}[h!]
\includegraphics[scale=0.65]{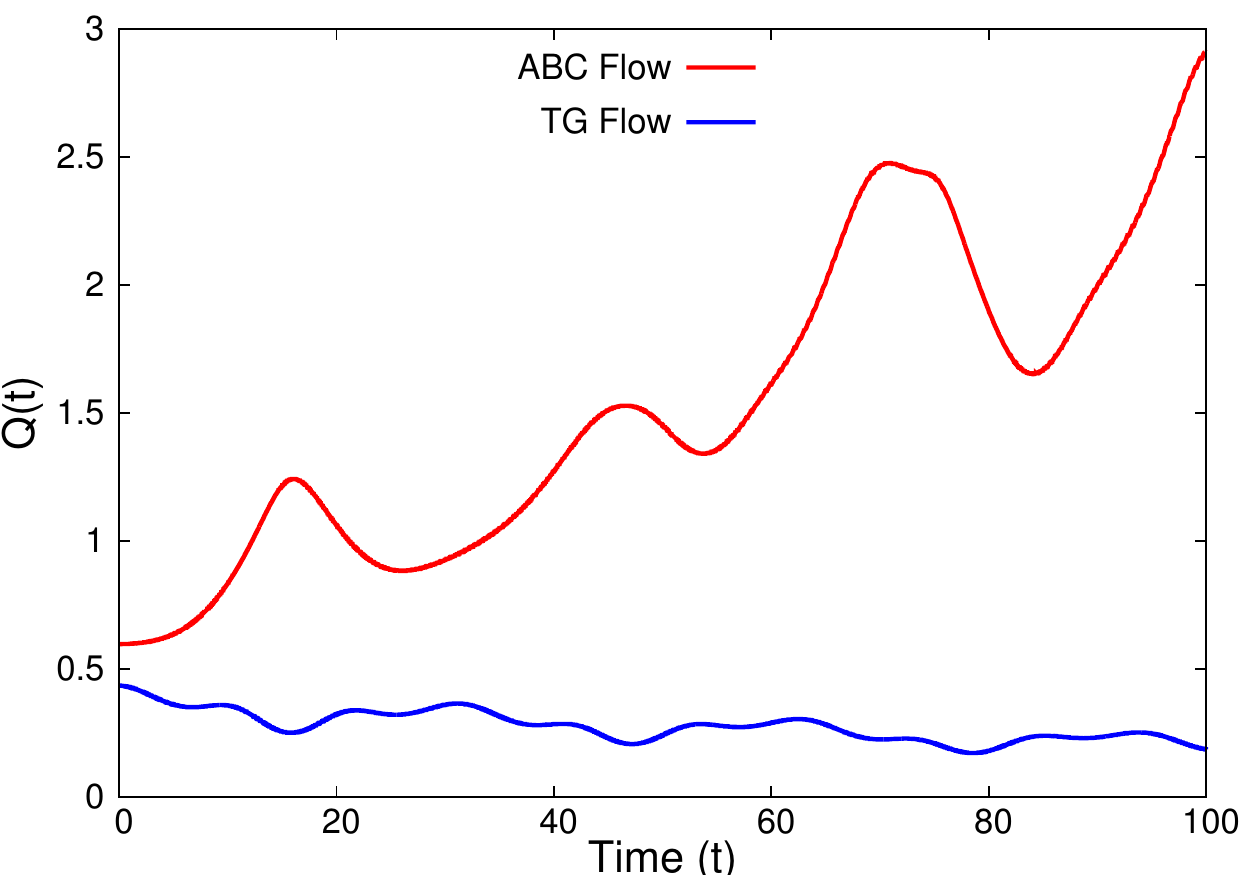}
\caption{(Color online) The time evolution of Rayleigh Quotient [Q(t)] for Taylor-Green (TG) and Arnold-Beltrami-Childress (ABC) flow.}
\label{Rayleigh}
\end{figure*}

We generalise the earlier definition of Rayleigh Quotient \cite{thyagaraja:1979} and use the following expression for our MHD system,
$$Q(t) = \frac{\int\limits_V \left[ \left( \vec{\nabla} \times \vec{u} \right)^2 + \frac{1}{2} \left( \vec{\nabla} \times \vec{B} \right)^2 \right] dV} {\int\limits_V \left( |\vec{u}|^2 + \frac{1}{2} |\vec{B}|^2 \right) dV} = \frac{\sum\limits_{k=0}^\infty k^2 |c_k|^2}{\sum\limits_{k=0}^\infty |c_k|^2}$$
where, $\vec{u}$ \& $\vec{B}$ are expanded in a Fourier series and $V$ is the volume of the three dimensional space. We evaluate this quantity for both the Taylor-Green and the Arnold-Beltrami-Childress flows and plot it as a function of time in Fig. \ref{Rayleigh}. From this figure it is seen that for the Taylor-Green flow the Rayleigh Quotient ($Q(t)$) remains bounded while for the Arnold-Beltrami-Childress flow the Rayleigh Quotient ($Q(t)$) increases as the time evolves. The occurrence of recurrence phenomenon in the TG flow is therefore consistent with earlier studies \cite{thyagaraja:1979,sen:2012} on hydrodynamic flows and other systems that have identified the boundedness of the Rayleigh quotient with the existence of recurrence. However an exact criterion determining the choice of initial conditions that can lead to recurrence remains an open issue. As noted by Thyagaraja \cite{thyagaraja:1981}  ``...the situation is more complicated in higher dimensions, and no general criterion is as yet available for deciding which initial data lead to non-collapsing recurrent motion''. \\


\section{Summary and conclusion}
\label{sec:summary}
We report the observation of recurrence of velocity and magnetic field isosurfaces for a three dimensional nearly ideal magnetohydrodynamic plasma. The choice of initial flow profile plays a crucial role in the recurrence phenomenon. Using Thyagaraja's theory \cite{thyagaraja:1979}, we observe that, only those flow profiles which keep the Rayleigh Quotient ($Q(t)$) bounded over time, are suitable candidates for the occurrence of recurrence. In our simulations we observe recurrence for an initial state that represents a Taylor-Green flow as seen from a periodic periodic reconstruction of the isosurfaces. Similar runs with an Arnold-Beltrami-Childress flow profile does not produce recurrence and the Rayleigh Quotient for this profile is also observed to grow boundlessly in time. \\

We have also carried out numerical simulation studies for two other flows viz. Roberts flow (Rayleigh quotient bounded) and Cats Eye flow (Rayleigh quotient not bounded) and found ``recurrence" to occur for Roberts flow and not happening for Cats Eye flow. However these results are not presented in this paper.\\

The presence of dissipation in our system due to finite viscosity and resistivity is seen to influence the long time behavior of recurrence due to dissipative decay of the initial state. Ideally true recurrence behavior in a dissipative system requires a driving source to compensate for the dissipation. We hope to carry out such a study in the future based on our present findings which clearly reveal recurrent behavior even in the absence of a driver. The other option for future studies is to seek recurrence behavior in conservative regularised versions \cite{thyagaraja:2010,thyagaraja:2014,govind:2016,govind:2018} of the three dimensional MHD systems which would be close to Hamilitonian system but be free of the numerical singularities associated with conservative systems. Apart from further numerical simulation and theoretical studies, it would be interesting to explore the possibilities of experimental observation of such recurrent phenomena in flows generated in laboratory plasma devices. The low dimensionality of the dynamics for certain flow patterns, as seen in our simulations, offers attractive potential options for predicting and controlling plasma behavior.



\section{Acknowledgement}
Simulations shown here are performed at Uday and Udbhav cluster at Institute for Plasma Research (IPR). RM thanks Samriddhi Sankar Ray at International Center for Theoretical Sciences, India for his initial help regarding pseudo spectral simulation and A Thyagaraja for his critical comments on the manuscript and his several helpful suggestions. The support from ICTS program: ICTS/Prog-dcs/2016/05 is also acknowledged. RM is grateful to Udaya Maurya at IPR for several numerical helps for three dimensional data visualisation used in this report. RM also acknowledges Govind Krishnaswami and Sonakshi Sachdev at Chennai Mathematical Institute for many helpful discussions.


\bibliography{biblio}

\begin{thebibliography}{10}

\bibitem{dauxois:2008}
Thierry Dauxois.
\newblock Fermi, pasta, ulam and a mysterious lady.
\newblock {\em arXiv preprint arXiv:0801.1590}, 2008.

\bibitem{zabusky:1965}
Norman~J Zabusky and Martin~D Kruskal.
\newblock Interaction of" solitons" in a collisionless plasma and the
  recurrence of initial states.
\newblock {\em Physical review letters}, 15(6):240, 1965.

\bibitem{chirikov:1979}
Boris~V Chirikov.
\newblock A universal instability of many-dimensional oscillator systems.
\newblock {\em Physics Reports}, 52(5):263 -- 379, 1979.

\bibitem{yuen:1978}
Henry~C Yuen and Warren~E Ferguson~Jr.
\newblock Fermi--pasta--ulam recurrence in the two-space dimensional nonlinear
  schr{\"o}dinger equation.
\newblock {\em The Physics of Fluids}, 21(11):2116--2118, 1978.

\bibitem{thyagaraja:1979}
A~Thyagaraja.
\newblock Recurrent motions in certain continuum dynamical systems.
\newblock {\em The Physics of Fluids}, 22(11):2093--2096, 1979.

\bibitem{landau:1959}
LD~Landau and EM~Lifshitz.
\newblock Fluid mechanics (chap. iii).
\newblock {\em Course of Theoretical Physics, Pergamon Press, London}, 6, 1959.

\bibitem{birkhoff:1960}
George~David Birkhoff.
\newblock {\em Dynamical systems}.
\newblock American mathematical society, 1960.

\bibitem{thyagaraja:1981}
A~Thyagaraja.
\newblock Recurrence, dimensionality, and lagrange stability of solutions of
  the nonlinear schr{\"o}dinger equation.
\newblock {\em The Physics of Fluids}, 24(11):1973--1975, 1981.

\bibitem{thyagaraja:1983}
A~Thyagaraja.
\newblock {\em Nonlinear Waves, Chapter 17}.
\newblock CUP Archive, 1983.

\bibitem{elgar:1990}
Elgar Steve, Freilich~M. H., and Guza~R. T.
\newblock Recurrence in truncated boussinesq models for nonlinear waves in
  shallow water.
\newblock {\em Journal of Geophysical Research: Oceans}, 95(C7):11547, 1990.

\bibitem{bryant:1988}
Peter~J Bryant.
\newblock Cyclic recurrence in nonlinear unidirectional ocean waves.
\newblock {\em Journal of Fluid Mechanics}, 192:329--337, 1988.

\bibitem{viswanath:2007}
Divakar Viswanath.
\newblock Recurrent motions within plane couette turbulence.
\newblock {\em Journal of Fluid Mechanics}, 580:339--358, 2007.

\bibitem{hogg:1982}
T~Hogg and BA~Huberman.
\newblock Recurrence phenomena in quantum dynamics.
\newblock {\em Physical Review Letters}, 48(11):711, 1982.

\bibitem{katok:1997}
Anatole Katok and Boris Hasselblatt.
\newblock {\em Introduction to the modern theory of dynamical systems},
  volume~54.
\newblock Cambridge university press, 1997.

\bibitem{tajima:1981}
T~Tajima, Martin~V Goldman, JN~Leboeuf, and JM~Dawson.
\newblock Breakup and reconstitution of langmuir wave packets.
\newblock {\em The Physics of Fluids}, 24(1):182--183, 1981.

\bibitem{kaw:1973}
Predhiman~Krishan Kaw, AT~Lin, and JM~Dawson.
\newblock Quasiresonant mode coupling of electron plasma waves.
\newblock {\em The Physics of Fluids}, 16(11):1967--1975, 1973.

\bibitem{vlad:1989}
G.~Vlad and A.~Bondeson.
\newblock Numerical simulations of sawteeth in tokamaks.
\newblock {\em Nuclear Fusion}, 29(7):1139, 1989.

\bibitem{thyagaraja:1999}
A~Thyagaraja, F.A. Haas, and D.J. Harvey.
\newblock A nonlinear dynamic model of relaxation oscillations in tokamak.
\newblock {\em Physics of Plasmas}, 6:2380, 1999.

\bibitem{pamela:2011}
S~J~P Pamela, G~T~A Huysmans, M~N~A Beurskens, S~Devaux, T~Eich, S~Benkadda,
  and JET~EFDA contributors.
\newblock Nonlinear mhd simulations of edge-localized-modes in jet.
\newblock {\em Plasma Physics and Controlled Fusion}, 53(5):054014, 2011.

\bibitem{jardin:2012}
SC~Jardin, N~Ferraro, J~Breslau, and J~Chen.
\newblock Multiple timescale calculations of sawteeth and other global
  macroscopic dynamics of tokamak plasmas.
\newblock {\em Computational Science \& Discovery}, 5(1):014002, 2012.

\bibitem{FFTW3:2005}
Matteo Frigo and Steven~G. Johnson.
\newblock The design and implementation of {FFTW3}.
\newblock {\em Proceedings of the IEEE}, 93(2):216--231, 2005.
\newblock Special issue on ``Program Generation, Optimization, and Platform
  Adaptation''.

\bibitem{ravi:2001}
Ravi Samtaney, Dale~I Pullin, and Branko Kosovi{\'c}.
\newblock Direct numerical simulation of decaying compressible turbulence and
  shocklet statistics.
\newblock {\em Physics of Fluids}, 13(5):1415--1430, 2001.

\bibitem{galloway:1986}
D~Galloway and Uriel Frisch.
\newblock Dynamo action in a family of flows with chaotic streamlines.
\newblock {\em Geophysical \& Astrophysical Fluid Dynamics}, 36(1):53--83,
  1986.

\bibitem{mayavi}
P.~Ramachandran and G.~Varoquaux.
\newblock {Mayavi: 3D Visualization of Scientific Data}.
\newblock {\em Computing in Science \& Engineering}, 13(2):40--51, 2011.

\bibitem{rupak:2018a}
Rupak Mukherjee, Akanksha Gupta, and Rajaraman Ganesh.
\newblock Compressibility effects on quasistationary vortex and transient hole
  patterns through vortex merger.
\newblock {\em arXiv preprint arXiv:1802.03240}, 2018.

\bibitem{rupak:2018b}
Rupak Mukherjee, Rajaraman Ganesh, Vinod Saini, Udaya Maurya, Nagavijayalakshmi
  Vydyanathan, and Bharatkumar Sharma.
\newblock Three dimensional pseudo-spectral compressible magnetohydrodynamic
  gpu code for astrophysical plasma simulation.
\newblock {\em arXiv preprint arXiv:1810.12707}, 2018.

\bibitem{rupak:2018c}
Rupak Mukherjee, Rajaraman Ganesh, and Abhijit Sen.
\newblock Coherent nonlinear oscillations in magnetohydrodynamic plasma at
  alfv$\backslash$'en resonance.
\newblock {\em arXiv preprint arXiv:1811.00744}, 2018.

\bibitem{bayly:1992}
BJ~Bayly, CD~Levermore, and T~Passot.
\newblock Density variations in weakly compressible flows.
\newblock {\em Physics of Fluids A: Fluid Dynamics}, 4(5):945--954, 1992.

\bibitem{terakado:2014}
Daiki Terakado and Yuji Hattori.
\newblock Density distribution in two-dimensional weakly compressible
  turbulence.
\newblock {\em Physics of Fluids}, 26(8):085105, 2014.

\bibitem{sen:2012}
Abhijit Sen, Dilip~P Ahalpara, Anantanarayanan Thyagaraja, and Govind~S
  Krishnaswami.
\newblock A kdv-like advection--dispersion equation with some remarkable
  properties.
\newblock {\em Communications in Nonlinear Science and Numerical Simulation},
  17(11):4115--4124, 2012.

\bibitem{thyagaraja:2010}
A~Thyagaraja.
\newblock Conservative regularization of ideal hydrodynamics and
  magnetohydrodynamics.
\newblock {\em Physics of Plasmas}, 17(3):032503, 2010.

\bibitem{thyagaraja:2014}
A~Thyagaraja.
\newblock Adjoint variational principles for regularised conservative systems.
\newblock In {\em AIP Conference Proceedings}, volume 1582, pages 107--115.
  AIP, 2014.

\bibitem{govind:2016}
Govind~S Krishnaswami, Sonakshi Sachdev, and Anantanarayanan Thyagaraja.
\newblock Local conservative regularizations of compressible
  magnetohydrodynamic and neutral flows.
\newblock {\em Physics of Plasmas}, 23(2):022308, 2016.

\bibitem{govind:2018}
Govind~S Krishnaswami, Sonakshi Sachdev, and Anantanarayanan Thyagaraja.
\newblock Conservative regularization of compressible dissipationless two-fluid
  plasmas.
\newblock {\em Physics of Plasmas}, 25(2):022306, 2018.

\end{thebibliography}


\end{document}